\documentclass[12pt]{article}
\usepackage{amsmath}
\usepackage{graphicx}
\usepackage{geometry}
\usepackage{authblk}
\usepackage{hyperref}
\usepackage{cite}

\title{Modeling the Effective Elastic Modulus and Thickness of Corrugated Boards Using Gaussian Process Regression and
Expected Hypervolume Improvement}

\author[1]{Ricardo Fitas\thanks{Corresponding Author: \href{mailto:ricardo.fitas@tu-darmstadt.de}{ricardo.fitas@tu-darmstadt.de}, ORCID: \href{https://orcid.org/0000-0001-5137-2451}{0000-0001-5137-2451}}}

\affil[1]{Technical University of Darmstadt, Alexanderstraße 8, 64283 Darmstadt}

\date{}

\begin{document}

\maketitle

\begin{abstract}
This work aims to model the hypersurface of the effective elastic modulus, \( E_{z, \text{eff}} \), and thickness, \( th_{\text{eff}} \), in corrugated boards. A Latin Hypercube Sampling (LHS) is followed by Gaussian Process Regression (GP), enhanced by EHVI as a multi-objective acquisition function. Accurate modeling of \( E_{z, \text{eff}} \) and \( th_{\text{eff}} \) is critical for optimizing the mechanical properties of corrugated materials in engineering applications. LHS provides an efficient and straightforward approach for an initial sampling of the input space; GP is expected to be able to adapt to the complexity of the response surfaces by incorporating both prediction and uncertainty. Therefore, the next points being generated and evaluated are based on the complexity of the hypersurfaces, and some points, especially those with higher variance, are more exploited and carry more importance. The performance of GP with EHVI is measured by Mean Squared Error (MSE). Prediction of GP resulted in \( \text{MSE}(E_{z, \text{eff}}) = 5.24 \, \text{kPa}^2 \) and \( \text{MSE}(th_{\text{eff}}) = 1 \, \text{mm}^2 \). GP possesses then improved accuracy and adaptability for future applications in structural optimization.

\end{abstract}


\section{Introduction}

A broad spectrum of industries — and especially the packaging sector — benefit from the unique attributes of corrugated boards, such as e-commerce, pharmaceutical, cosmetics, food, and beverage industries \cite{jestratijevic2022sustainable, enlund2021sustainable}. Their high strength-to-weight ratio, cost-effectiveness, and versatility provide an ideal combination of qualities for protecting and displaying products. In 2021, the market dynamics demonstrated the growing significance of this industry, which is anticipated to be worth USD 199.8 billion and is expected to grow at a Compound Annual Growth Rate (CAGR) of 5.0\% to reach a stunning USD 254.5 billion by 2026 \cite{jestratijevic2022sustainable, paper_paperboard_packaging_market_2021, global_packaging_market_2022}. The fluting medium creates a structure that is consistently superior to solid fiberboards, owing to the board's excellent mechanical performance, including resistance to compression and bending forces.

A major problem in corrugated board design is related to the determination of the effective elastic modulus. It allows to predict the stiffness of the board when loaded. Stakeholders in the design and manufacturing of corrugated boards need a straightforward and fast-to-compute function that allows them to compute the effective elastic modulus based on their board design \cite{su152115588}. That could lead to decision-making and designs that are both more informed and faster, also arguably reducing the volume of materials needed to manufacture a given board.

While homogenization methods currently exist for calculating the effective elastic modulus of corrugated boards \cite{starke2020material, marek2017homogenization}, these methods are computationally demanding, especially when incorporated into optimization tasks that require multiple iterations. These calculations take a lot of time and pose a significant barrier to the ability to optimize in real-time. They limit the ability to explore many different design options and to do so efficiently.

This task can be resolved through the creation of a surrogate model that preserves the results of the homogenization method as much as possible insofar as it dramatically cuts computation time. There is a need for a model that allows the rapid iterative calculation necessary for performing optimization operations while still maintaining the integrity of the solution.

This work presents a surrogated model to tackle such research gaps in the literature. The approach is based on a combination of Latin Hypercube Sampling (LHS) to generate the data and Gaussian Process (GP) regression to develop an efficient surrogate model that is expected to improve the computational time.

The remainder of the paper is structured as follows: Section 2 details the methodology, including the homogenization process, data generation, and the development of the surrogate model. Section 3 presents the results of the applied methodology. Section 4 provides conclusions and future work directions.

\section{Methodology}

\subsection{Homogenization Method}

\begin{figure}[ht]
    \centering
    \includegraphics[width=\textwidth]{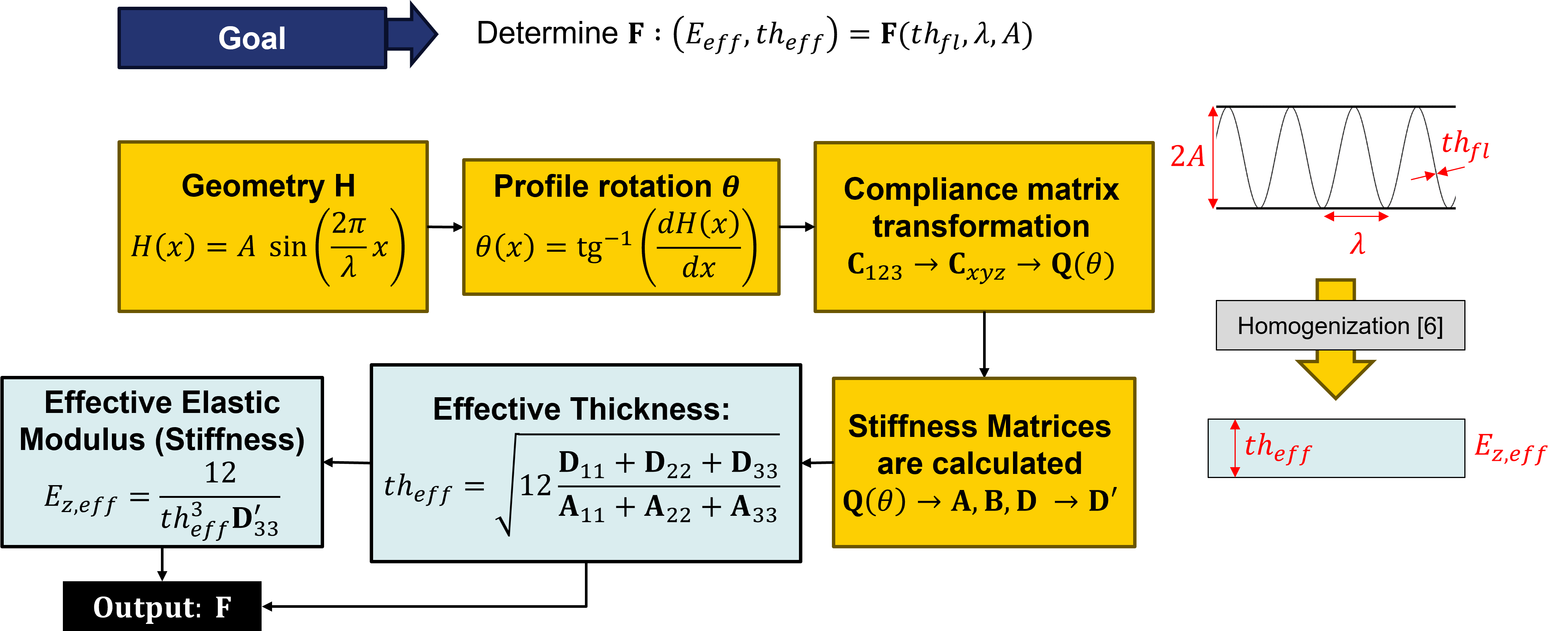}
    \caption{Workflow to compute effective stiffness and thickness of corrugated board structures.}
    \label{fig0}
\end{figure}

The homogenization process is used to compute the effective elastic modulus ($E_{z, \text{eff}}$) and thickness ($th_{\text{eff}}$) of corrugated boards by considering the geometric and material properties of the flute structure. In this study, the process follows several key steps, including defining the material compliance matrix and performing transformations to account for the flute's geometry. The flowchart in Figure \ref{fig0} provides a representation of the homogenization method used.

\subsubsection{Material Compliance Matrix}

The compliance matrix for the corrugated board is defined based on the elastic properties of the material, including the Young's moduli ($E_1$, $E_2$, and $E_3$), the shear moduli ($G_{12}$, $G_{13}$, and $G_{23}$), and the Poisson ratios ($\nu_{12}$, $\nu_{13}$, and $\nu_{23}$). The general form of the compliance matrix ($C_{123}$) in the material coordinate system is given by:

\[
C_{123} = \begin{bmatrix}
    \frac{1}{E_1} & -\frac{\nu_{12}}{E_1} & -\frac{\nu_{13}}{E_1} & 0 & 0 & 0 \\
    -\frac{\nu_{12}}{E_1} & \frac{1}{E_2} & -\frac{\nu_{23}}{E_2} & 0 & 0 & 0 \\
    -\frac{\nu_{13}}{E_1} & -\frac{\nu_{23}}{E_2} & \frac{1}{E_3} & 0 & 0 & 0 \\
    0 & 0 & 0 & \frac{1}{G_{12}} & 0 & 0 \\
    0 & 0 & 0 & 0 & \frac{1}{G_{13}} & 0 \\
    0 & 0 & 0 & 0 & 0 & \frac{1}{G_{23}}
\end{bmatrix}
\]

\subsubsection{Flute Geometry and Transformation}

The corrugated structure is modeled as a periodic sinusoidal profile, described by:

\[
H(x) = \frac{f_{\text{start}}}{2} \sin \left( \frac{2 \pi}{p} x \right)
\]

where $f_{\text{start}}$ is the initial flute height, and $p$ is the wavelength of the flute. The profile's rotation angle $\theta$ at any point along the x-axis is given by:

\[
\theta = \tan^{-1} \left( \frac{\pi f_{\text{start}}}{p} \cos \left( \frac{2 \pi}{p} x \right) \right)
\]

To account for the periodic nature of the flutes, transformation matrices are applied to the compliance matrix. The transformation matrices for rotation about the y-axis are defined as:

\[
T_e = \begin{bmatrix}
    c^2 & 0 & s^2 & 0 & -sc & 0 \\
    0 & 1 & 0 & 0 & 0 & 0 \\
    s^2 & 0 & c^2 & 0 & sc & 0 \\
    0 & 0 & 0 & c & 0 & -s \\
    2sc & 0 & -2sc & 0 & c^2 - s^2 & 0 \\
    0 & 0 & 0 & -s & 0 & c
\end{bmatrix}
\]

where $c = \cos(\theta)$ and $s = \sin(\theta)$. The transformed compliance matrix in the global coordinate system ($C_{xyz}$) is calculated as:

\[
C_{xyz} = T_e C_{123} T_s
\]

where $T_s$ is the transpose of $T_e$.

\subsubsection{Homogenization Along the Thickness}

The homogenization method involves integrating the material properties through the thickness of the fluting structure. The vertical thickness of the fluting, $t_v$, is calculated as:

\[
t_v = \frac{t}{\cos(\theta)}
\]

where $t$ is the thickness of the liner, the effective elastic modulus and effective thickness are computed by integrating the stiffness matrix components ($A$, $B$, $D$, and $F$ matrices) over the flute's periodic length $p$. The integral for $A_{11}$, for example, is given by:

\[
A_{11} = \frac{1}{p} \int_0^p t_v Q_{11} \, dx
\]

where $Q_{11}$ is a reduced stiffness component derived from the transformed compliance matrix.

\subsubsection{Effective Thickness and Elastic Modulus}

The final step in the homogenization process is calculating the effective thickness ($th_{\text{eff}}$) and the effective elastic modulus ($E_{z, \text{eff}}$). The effective thickness is calculated using:

\[
th_{\text{eff}} = \sqrt{\frac{12 \left( D_{11} + D_{22} + D_{33} \right)}{A_{11} + A_{22} + A_{33}}}
\]

The effective elastic modulus is then determined by:

\[
E_{z, \text{eff}} = \frac{12}{t_h^3 D^{-1}_{33}}
\]

This homogenization process allows us to predict the effective mechanical properties of the corrugated board, considering the geometry and material characteristics, enabling optimization of the board's design.

\subsection{Data Generation}

In order to build a surrogate model capable of predicting the effective elastic modulus ($E_{z, \text{eff}}$) and effective thickness ($th_{\text{eff}}$) of corrugated boards, a dataset is generated using a combination of Latin Hypercube Sampling (LHS) and Gaussian Process (GP) regression. The variables (features) for this process are the flute thickness, flute height (or amplitude), and wavelength, while the outputs (targets) are $E_{z, \text{eff}}$ and $th_{\text{eff}}$.

The data generation process is divided into two parts: \begin{itemize} \item \textbf{Initial Sampling:} 20\% of the samples are generated using Latin Hypercube Sampling (LHS) to ensure that the initial set of points uniformly explores the input space. \item \textbf{Active Learning with GP:} 80\% of the data is generated iteratively using GP regression, leveraging an acquisition function to sample points that improve the model based on Expected Improvement (EI). \end{itemize}

\subsubsection{Initial Sampling}

First, the initial data points are sampled using LHS. The ranges for the input variables are: 

\begin{align*} 
\text{Flute thickness} &\in [5 \times 10^{-3}, 2 \times 10^{-2}] , \text{m}, \\
\text{Amplitude} &\in [1 \times 10^{-3}, 1 \times 10^{-2}] , \text{m}, \\ \text{Wavelength} &\in [1 \times 10^{-4}, 2 \times 10^{-3}] , \text{m}. \\
\end{align*}

These initial data points are used to evaluate the outputs $E_{z, \text{eff}}$ and $th_{\text{eff}}$, which are computed using the homogenization process described earlier.

\subsubsection{Active Learning Loop}

Once the initial GP models are trained, an active learning loop is employed to iteratively improve the model. At each iteration, 1000 candidate points are generated, and the GP model predicts both the mean and uncertainty for $E_{z, \text{eff}}$ and $th_{\text{eff}}$ for these candidates.

The Expected Improvement (EI) acquisition function is then used to identify the next point to sample. The EI function is designed to select points that are expected to yield the greatest improvement in the model’s predictive capability:
\[
\text{EI}(X) = \sigma(X) \left( z \Phi(z) + \phi(z) \right)
\]
where
\[
z = \frac{y^* - \mu(X)}{\sigma(X)}
\]

and $y^*$ is the current best value observed, $\mu(X)$ and $\sigma(X)$ are the mean and standard deviation predicted by the GP model at point $X$, $\Phi(z)$ is the cumulative distribution function (CDF), and $\phi(z)$ is the probability density function (PDF) of the standard normal distribution.

The point that maximizes the EI function is selected, and the true values of $E_{z, \text{eff}}$ and $th_{\text{eff}}$ at this point are computed using the homogenization method.

The final dataset includes 1000 data points: 200 from the initial LHS sampling and 800 from the GP-based active learning process.

\section{Results}

Table \ref{tab0} presents the paper material properties used to run the experiments.

\begin{table}[ht]
\centering
\caption{Material Properties of Paper} \label{tab0}
\begin{tabular}{|l|l|l|}
\hline
\textbf{Property} & \textbf{Symbol} & \textbf{Value} \\ \hline
Young's modulus MD & \( E_1 \) & \( 1.709 \times 10^9 \, \text{Pa} \) \\ \hline
Young's modulus CD & \( E_2 \) & \( 0.918 \times 10^9 \, \text{Pa} \) \\ \hline
Young's modulus ZD & \( E_3 \) & \( E_1 / 190 = 8.995 \times 10^6 \, \text{Pa} \) \\ \hline
Shear modulus in MD-CD plane & \( G_{12} \) & \( 0.387 \sqrt{E_1 E_2} \approx 0.485 \times 10^9 \, \text{Pa} \) \\ \hline
Poisson's ratio in MD-CD plane & \( \nu_{12} \) & \( 0.293 \sqrt{E_1 / E_2} \approx 0.4 \) \\ \hline
Shear modulus in MD-ZD plane & \( G_{13} \) & \( E_1 / 55 = 3.107 \times 10^7 \, \text{Pa} \) \\ \hline
Shear modulus in CD-ZD plane & \( G_{23} \) & \( E_2 / 35 = 2.623 \times 10^7 \, \text{Pa} \) \\ \hline
Poisson's ratio in MD-ZD plane & \( \nu_{13} \) & \( 0.001 \) \\ \hline
Poisson's ratio in CD-ZD plane & \( \nu_{23} \) & \( 0.001 \) \\ \hline
\end{tabular}
\end{table}

\begin{figure}[ht]
    \centering
    \includegraphics[width=0.6\textwidth]{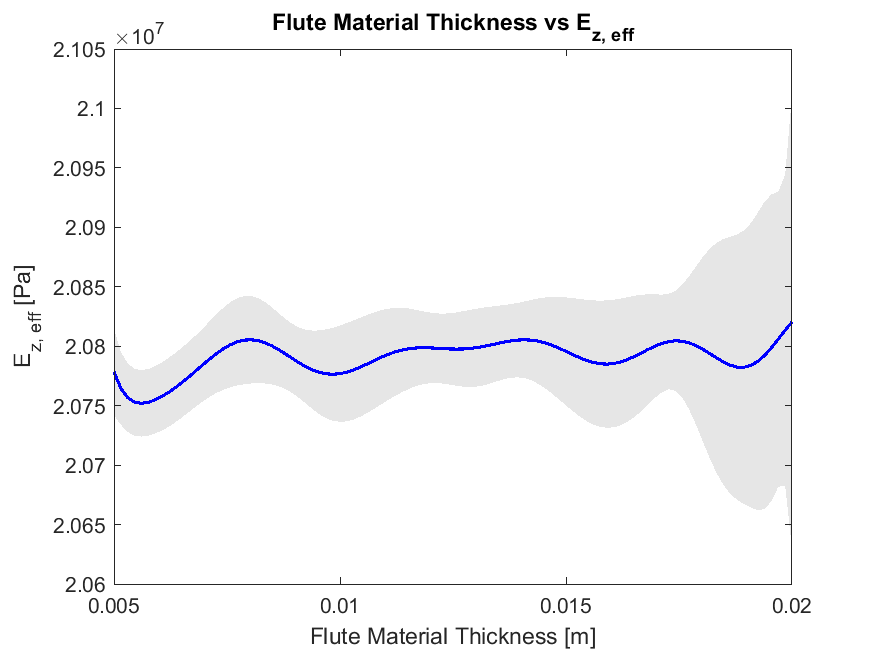}
    \caption{Effective stiffness as a function of the Flute Material Thickness}
    \label{fig22}
\end{figure}

\begin{figure}[ht]
    \centering
    \includegraphics[width=0.6\textwidth]{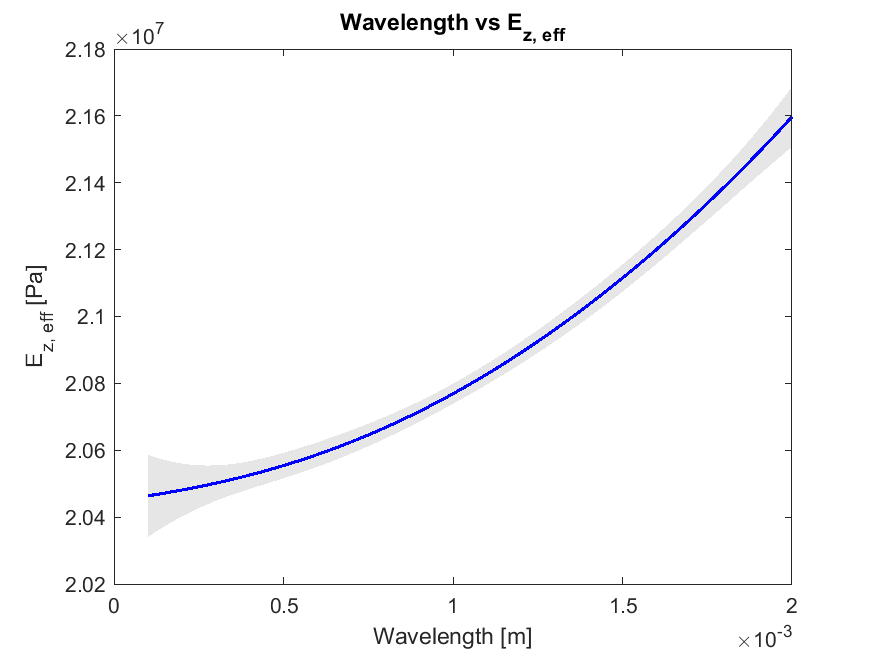}
    \caption{Effective stiffness as a function of Wavelength}
    \label{fig3}
\end{figure}

\begin{figure}[ht]
    \centering
    \includegraphics[width=0.6\textwidth]{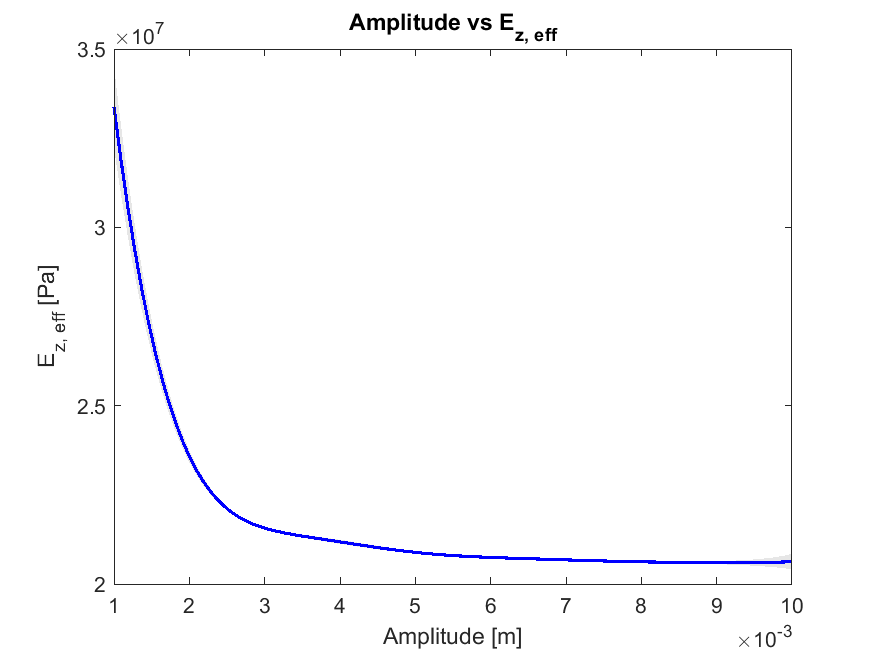}
    \caption{Effective stiffness as a function of the Amplitude}
    \label{fig4}
\end{figure}

\begin{figure}[ht]
    \centering
    \includegraphics[width=0.6\textwidth]{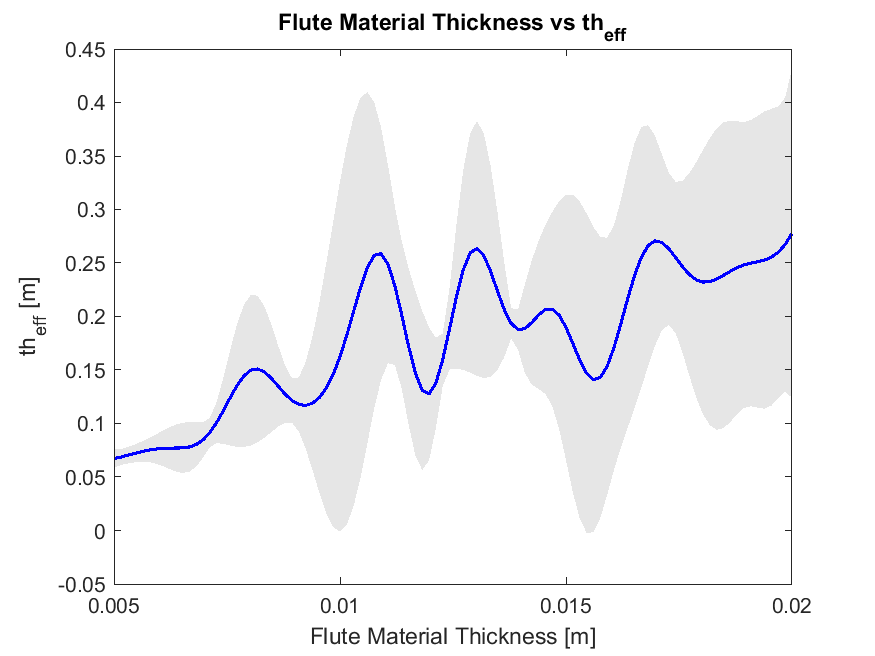}
    \caption{Effective thickness as a function of the Flute Material Thickness}
    \label{fig5}
\end{figure}

\begin{figure}[ht]
    \centering
    \includegraphics[width=0.6\textwidth]{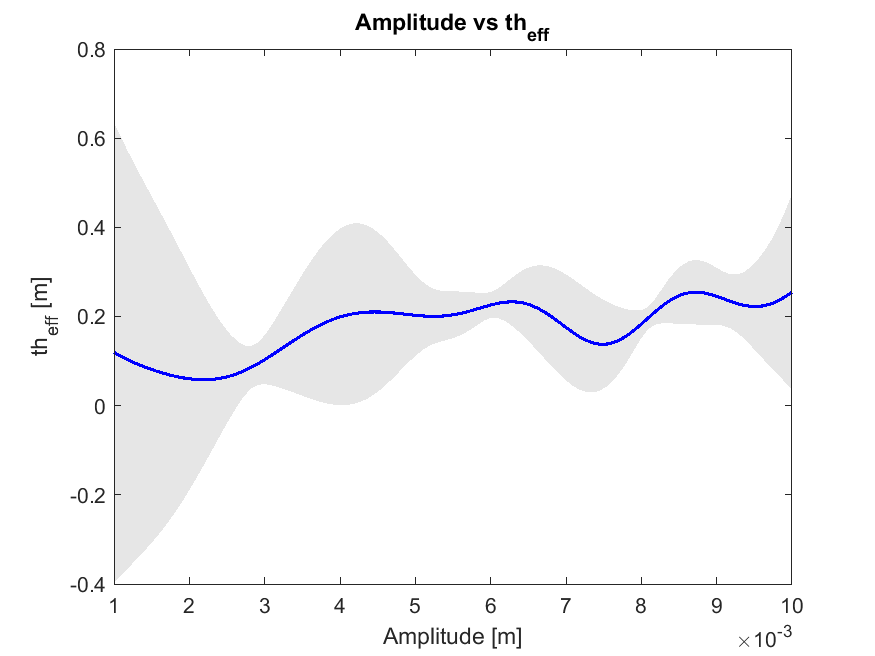}
    \caption{Effective thickness as a function of the Amplitude}
    \label{fig6}
\end{figure}

\begin{figure}[ht]
    \centering
    \includegraphics[width=0.6\textwidth]{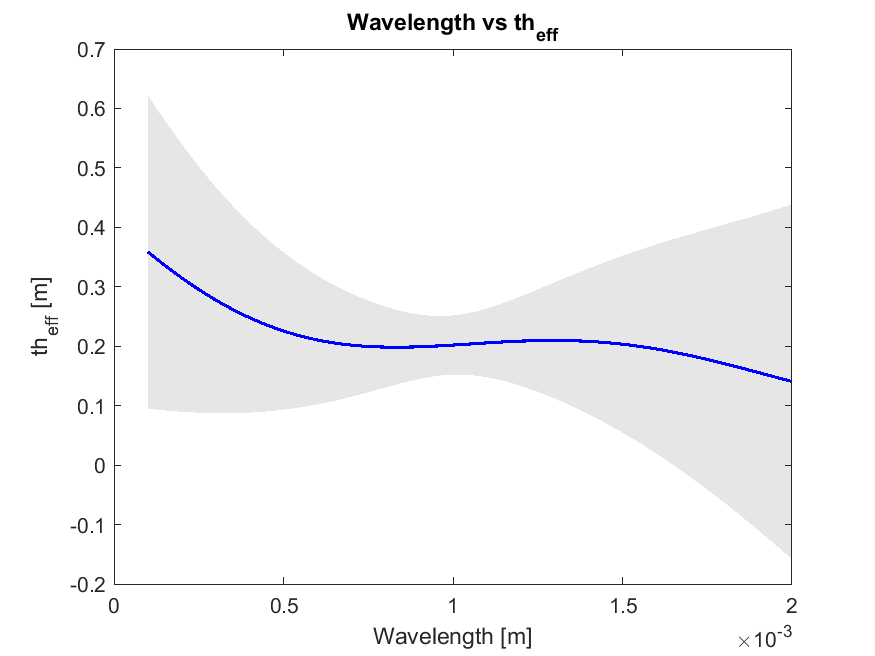}
    \caption{Effective thickness as a function of Wavelength}
    \label{fig7}
\end{figure}

In the Figures \ref{fig22} to \ref{fig7}, the effects of different features, namely flute material thickness, wavelength, and amplitude, on the two dependent variables — effective thickness (\( t_{\text{eff}} \)) and effective stiffness (\( E_{z,\text{eff}} \)) — are analyzed. One can evaluate the projected values as well as the confidence intervals, which are represented as shaded areas on the plots, by predicting the mean response and uncertainty bounds.

As shown in Figures \ref{fig22}, \ref{fig3}, and \ref{fig4}, the features that have the greatest effects on the effective stiffness are the wavelength and amplitude. While the amplitude exhibits a rapid decline in stiffness as it increases, showing a nonlinear impact, the wavelength exhibits a significant positive correlation with the stiffness, suggesting that higher wavelengths lead to enhanced \( E_{z,\text{eff}} \). The flute material thickness seems to be the main influencing factor for effective thickness, as shown in Figure \ref{fig5} causing noticeable changes in \( th_{\text{eff}} \) as it varies. These relationships are well captured by the GP model, which also emphasizes how sensitive \( t_{\text{eff}} \) is to changes in flute material thickness.

Regarding uncertainty, each plot's shaded areas show that it is comparatively low for amplitude and wavelength in the prediction of the effective stiffness, whereas it is relatively high in the other pairs.

A statistical anaylsis of the performance of the proposed model is here done, especifically when comparing GP method with LHS alone. The hypothesis posits that LHS method alone captures fewer model complexities than the combined LHS and GP approach. This assumption is evaluated by analyzing the disparities in power function exponents, which measure the magnitude of response variation in the model outputs. Two types of simulations on the generated points were performed: one estimating the curvature and another estimating the magnitude. The complexity captured by each method is inferred from the exponent magnitudes of fitted power functions, with higher values indicating greater complexity.

The points are randomly sorted from the datasets. The gradient magnitude of \( E_{z, \text{eff}} \) and \( th_{\text{eff}} \) can be computed as follows:

\[
|\nabla E_{z, \text{eff}}|(i) = \left\| \frac{E_{z, \text{eff}}(i+1) - E_{z, \text{eff}}(i-1)}{\| \mathbf{X}(i+1) - \mathbf{X}(i-1) \|} \right\|
\]
\[
|\nabla th_{\text{eff}}|(i) = \left\| \frac{th_{\text{eff}}(i+1) - th_{\text{eff}}(i-1)}{\| \mathbf{X}(i+1) - \mathbf{X}(i-1) \|} \right\|
\]

where \( \mathbf{X}(i) \) represents the feature vector from matrix  \( \mathbf{X} \) at index \( i \).

The curvature of \( E_{z, \text{eff}} \) and \( th_{\text{eff}} \) (using a Hessian approximation) can be computed as:

\[
\kappa_{E_{z, \text{eff}}}(i) = \left\| \frac{E_{z, \text{eff}}(i+1) - 2E_{z, \text{eff}}(i) + E_{z, \text{eff}}(i-1)}{\| \mathbf{X}(i+1) - \mathbf{X}(i) \|^2} \right\|
\]
\[
\kappa_{th_{\text{eff}}}(i) = \left\| \frac{th_{\text{eff}}(i+1) - 2th_{\text{eff}}(i) + th_{\text{eff}}(i-1)}{\| \mathbf{X}(i+1) - \mathbf{X}(i) \|^2} \right\|
\]

where \( \| \cdot \| \) denotes the Euclidean norm.

Model 1 (100\% LHS) and Model 2 (20\% LHS + 80\% GP) are compared in Tables  \ref{tab1} and \ref{tab2} using these performance metrics. Model 2 exhibits significantly higher exponent magnitudes, confirmed by a Likelihood Ratio Test (LRT) with a p-value $< 0.001$, supporting the hypothesis that combining LHS with GP increases complexity capture.

\begin{table}[ht]
\centering
\caption{Results: LHS vs GP} \label{tab1}
\resizebox{\textwidth}{!}{%
\begin{tabular}{|c|c|c|}
\hline
\textbf{Metric} & \textbf{LHS (100\%) [Model 1]} & \textbf{LHS (20\%) + GP (80\%) [Model 2]} \\
\hline
Mean Squared Error (MSE) & N/A & $E_{z,\text{eff}}: 5.24 \, [kPa^2]$ \\
\hline
Power Function (curvature) & $E_{z,\text{eff}}: y = 349x^{0.0022}$ & $E_{z,\text{eff}}: y = 19x^{0.1182}$ \\
 & $t_{h,\text{eff}}: y = 378x^{-0.0027}$ & $t_{h,\text{eff}}: y = 314x^{0.0176}$ \\
\hline
Power Function (magnitude) & $E_{z,\text{eff}}: y = 362x^{0.0009}$ & $E_{z,\text{eff}}: y = 525x^{-0.0195}$ \\
 & $t_{h,\text{eff}}: y = 342x^{0.0233}$ & $t_{h,\text{eff}}: y = 342x^{-0.3050}$ \\
\hline
\end{tabular}%
}
\label{table:results_lhs_vs_gp}
\end{table}

\begin{table}
\centering
\caption{Tail Index via MLE (Pareto Distribution) for LHS Method}  \label{tab2}
\begin{tabular}{|c|c|c|}
\hline
\textbf{Parameter} & \textbf{Tail Index (Model 1)} & \textbf{Tail Index (Model 2)}\\
\hline
A & 0.972611 & 0.231142 \\
\hline
B & 0.505299 & 0.387527 \\
\hline
C & 1.046491 & 1.060603 \\
\hline
D & 0.846324 & 1.058890 \\
\hline
\end{tabular}
\label{table:tail_index_lhs}
\end{table}

The Q-Q plots (Figures 8 and 9) provide a visualization of tail heaviness. Model 1, with only LHS, demonstrates some heavy tails, though most values align with the 45-degree line, indicating a challenge in capturing full model complexity. In contrast, Model 2 (LHS + GP) shows heavier tails, especially for \( t_{h, \text{eff}} \), underscoring the increased complexity captured by incorporating GP.

\begin{figure}
    \centering
    \includegraphics[width=\textwidth]{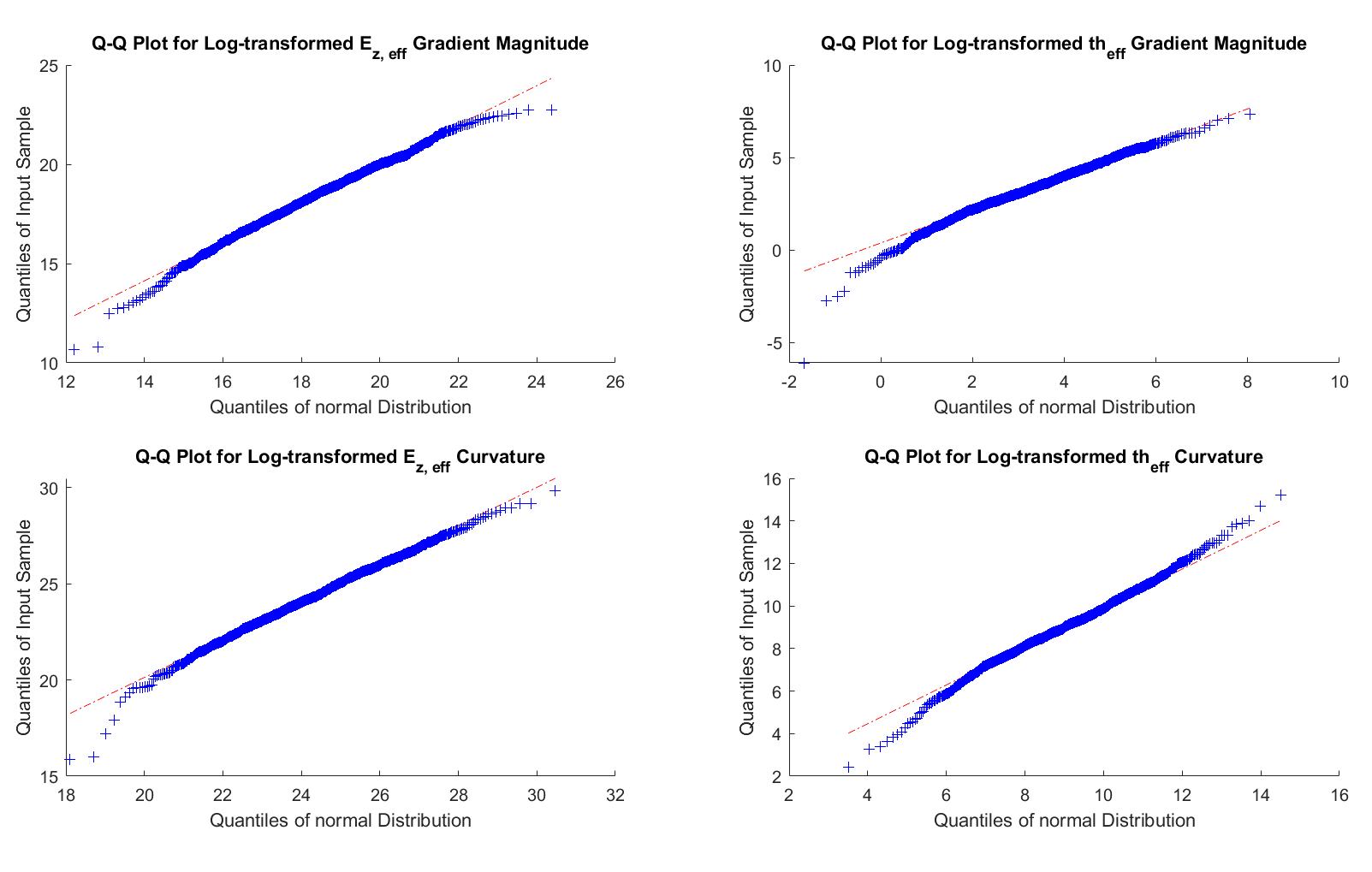}
    \caption{Q-Q Plots for Log-transformed effective stiffness and thickness: Latin Hypercube Sampling}
    \label{fig8}
\end{figure}

\begin{figure}
    \centering
    \includegraphics[width=\textwidth]{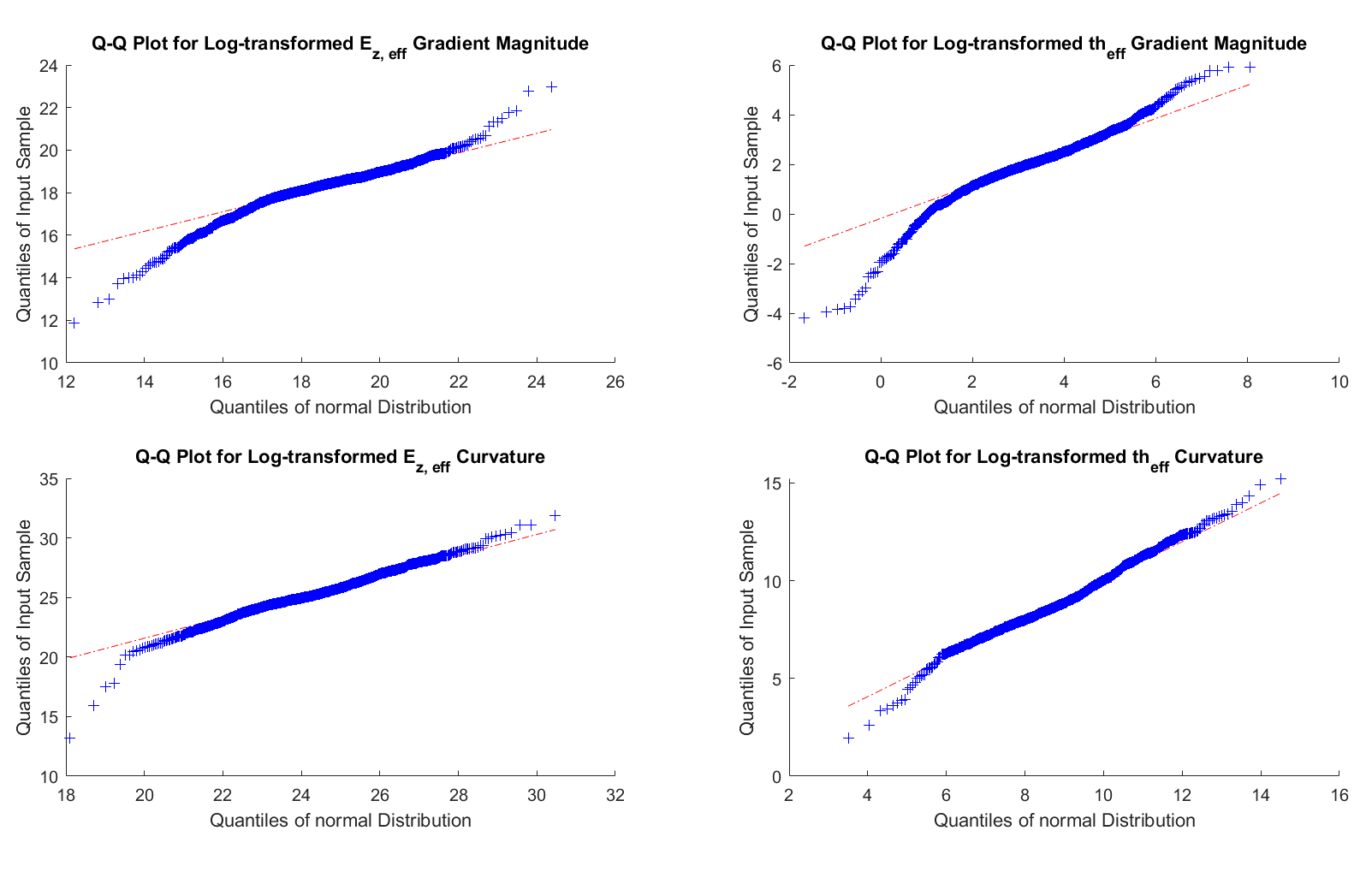}
    \caption{Q-Q Plots for Log-transformed effective stiffness and thickness: Gaussian Process Regression}
    \label{fig9}
\end{figure}

The Tail Index analysis (Table \ref{tab2}) using Maximum Likelihood Estimation (MLE) for Pareto distribution fitting reveals no significant difference in tail heaviness for \( t_{h, \text{eff}} \), but Model 2 displays increased tail indices in most parameters, aligning with the hypothesis that GP captures additional complexities.

\section{Conclusions}

This study evaluated the effectiveness of combining Latin Hypercube Sampling (LHS) with Gaussian Process (GP) models in capturing model complexities for predicting effective elastic modulus and thickness. The results affirm that incorporating GP leads to heavier tails in the distribution, as observed in this model (Model 2). This characteristic is crucial for model reliability, indicating the model’s capacity to detect subtle patterns within the data.

However, increasing complexity through GP introduces a trade-off, as it potentially raises sensitivity to outliers, making the model more prone to overfitting in edge cases. For future research, alternative architectures such as Random Forests should be explored post-GP to balance accuracy and robustness. Further validation across different datasets and scenarios will enhance the model’s effectiveness in predicting material behavior under varied conditions. Moreover, a better estimation of the variation should be done as well, by implementing a different independent GP model.

\vspace{12pt} 
\noindent \textbf{\small Acknowledgements:} 
The author gratefully acknowledges the support provided by the Chair of Paper Technology and Mechanical Process Engineering at TU Darmstadt.

\bibliographystyle{unsrt} 
\bibliography{ref1}

\end{document}